# Measurements of W-value, Mobility and Gas Gain in Electronegative Gaseous $CS_2$ and $CS_2$ Gas Mixtures

**Kirill Pushkin and Daniel Snowden-Ifft**

Department of Physics, Occidental College, Los Angeles, CA 90041, USA

**Corresponding Author: Kirill Pushkin**

**Affiliation: Occidental College**

**Postal Address: 1600 Campus Rd., Los Angeles, CA 90041, USA**

**Telephone: 1-323-259-2569**

**E-mail: kpushkin@oxy.edu**

## Abstract

W-value, mobility and gas gain measurements have been carried out in electronegative gaseous $CS_2$ and $CS_2$ gas mixtures at a pressure of 40 Torr making use of a single electron proportional counter method. The experimental results have revealed that W-values obtained for $CS_2$ (40 Torr), $CS_2$-$CF_4$ (30 Torr - 10 Torr), $CS_2$-Ar (35 Torr - 5 Torr), $CS_2$-Ne (35 Torr - 5 Torr) and $CS_2$-He (35 Torr - 5 Torr) gas mixtures are 21.1±2.7(stat)±3(syst) eV, 16.4±1.8(stat)±2(syst) eV, 13.1±1.5(stat)±2(syst) eV, 16.3±3.0(stat)±3(syst) eV and 17.3±3.0(stat)±3(syst) eV. The mobility for all $CS_2$ gas mixtures was found to be slightly greater and the gas gain was found to be significantly greater relative to pure $CS_2$.

## 1. Introduction

Time Projection Chambers (TPCs) have been widely used for the last few decades. Image transport in TPCs is typically implemented with electrons with high drift velocities, ~1000 m/s, and large diffusion [1,2]. In the Negative Ion TPC (NITPC) concept, which has recently been proposed and developed [3], electrons are captured by electronegative gases forming negative ions whose diffusion is substantially reduced. It has been shown [4] that diffusion of negative ions in pure, electronegative $CS_2$ reduces to thermal levels providing high spatial resolution in the detector though with very slow transport of negative ions ~50m/s. The low diffusion in a NITPC is crucial in, for example, the search for Dark Matter in the Galaxy [5] or the search for neutrinoless double beta decay where precise reconstruction of the events (3D tracks) is needed [6].

DRIFT (Directional Recoil Identification From Tracks) detector based on electronegative, low-pressure carbon disulfide gas ($CS_2$) was the first NITPC proposed and used to search for Dark Matter in the Galaxy [3-5]. This detector consists of two back-to-back TPCs each with Multi Wire Proportional counters (MWPCs) readout providing three measurements of the range components of the tracks (Δx, Δy, Δz). When electrons are captured on electronegative $CS_2$, negative molecules are drifted to the anode wires of the MWPCs where the high electric field there strips off the attached electrons, which then produce a normal electron avalanche. The W-value, the average energy lost by the incident particle per ion pair, negative ion mobility and gas gain play an important role in the DRIFT experiment. For instance, measurements of W-value provide a precise estimation of ionization yield from 5.9 keV X-rays emitted from the calibration source $^{55}$Fe and allow DRIFT to calculate the ionization yield from nuclei recoils [7]. Mobility values allow the time profile of event signals to be converted into measurements of the component of the track perpendicular to the MWPCs. Gas gain measurements allow for a quantitative comparison of various gas mixtures with an eye towards improving signal to electronic noise. The ability of the DRIFT detector to run with different gas mixtures, already demonstrated with Ar and Xe, is a tremendous advantage [4,8]. Carbon tetrafluoride ($CF_4$) is a prominent "cool" molecular gas [9,10] and a spin-dependent candidate for Weakly Interactive Massive Particles (WIMPs) searches [11]. Its scintillating properties along with some noble gases Ar, Ne, and He are of an interest for future research with regards to providing a start signal for events in the fiducial volume of a NITPC. In this work we report on measurements of W-value, mobility and gas gain in $CS_2$, $CS_2$-$CF_4$ $CS_2$-Ar, $CS_2$-Ne, $CS_2$-He gaseous mixtures.

## 2. W-value measurements

*A.* Introduction

The method of W-value measurements with proportional counters was developed by Srdoc et al., [12-15] and used to measure W-values both in noble and molecular gases. It relies on the measurements of the average pulse heights produced by single electrons (SE) and by low energy X-rays under the same experimental conditions. In this method, the ionization yield, $N_{ion}$, is calculated as a ratio of a pulse height parameter $X_{^{55}Fe}$, which

68   is proportional to the total charge deposited on an anode wire from 5.9 keV X-rays
69   emitted from an $^{55}$Fe source to a similar pulse height parameter $X_{SE}$ from single electrons
70   (SE):

$$N_{ion} = \frac{X_{^{55}Fe}}{X_{SE}} \times \frac{G_{SE}}{G_{^{55}Fe}} \quad (1)$$

72   where $G_{SE}$ and $G_{^{55}Fe}$ are the electronic gains used in measuring X-rays from the $^{55}$Fe
73   source and SE events.
74       Having obtained the ionization yield the W–value can be found using:

$$W = \frac{E_{^{55}Fe}}{N_{ion}} \quad (2)$$

76
77   *B*. The Detector
78       A proportional tube has been constructed to measure the W-value, mobility and gas
79   gain in pure $CS_2$ and $CS_2$ gas mixtures. The proportional counter, made out of copper,
80   was 58 cm long with a diameter of 2.54 cm and a slit of about 0.6 cm, which ran along
81   most of the length of the tube and allowed radiation to pass into the proportional counter.
82   The central anode wire was 100 μm gold coated tungsten. All measurements were made
83   at -1600 V and -1550 V applied to the copper tube in pure $CS_2$ and $CS_2$ gas mixtures,
84   respectively, while the anode wire was at ground. The proportional counter was placed
85   within a stainless steel chamber with an interior volume of ~270 l and filled with 40 Torr
86   of $CS_2$ or $CS_2$ gas mixtures. The $^{55}$Fe source was placed in a holder with an automatic
87   shutter allowing operation of the source inside of the chamber during the measurements.

All electrical cables were connected to the detector and the source holder through BNC feedthroughs mounted on the stainless steel chamber.

### C. Measurements of 5.9 keV X-rays from an $^{55}$Fe source.

A schematic of the experimental set-up and electronics for the measurements of 5.9 keV X-rays from the $^{55}$Fe source is shown in Fig.1. All data were taken using standard NIM electronics for trigger logic and event collection. As seen from the schematic, the anode charge was collected with a charge sensitive preamplifier (Amptek-250) with a time constant of 300 μs. All signals were then sent to an amplifier (Ortec-855) with a shaping time constant of 3 μs and gain of 5. Triggers were formed from the signals and then recorded to a data acquisition (DAQ) computer. The rate of X-ray interactions in the proportional counter during the measurements was of the order of 50 counts/s to avoid pile-up of the events. The background data were taken and eventually subtracted from the $^{55}$Fe events. The charge sensitive preamplifier was calibrated with a pulse generator and its drift was less than 1%.

### D. Measurements of single electron events

The experimental set-up for the measurements of SE events is shown in Fig.2. A xenon flashlamp (EG&G Optoelectronics, LS-1102) generated UV photons with high efficiency from 180 nm (air-cutoff) to 400 nm. A small hole was drilled in the side of the xenon flashlamp cover allowing a photodiode (PD) to see the flashes. The xenon flashlamp was operated with a pulse function generator at a frequency of 100Hz. The UV photons traveled through several pieces of ~400 μm thick transparent plastic used for attenuation, a 200 μm hole in a piece of aluminum for further attenuation and a sapphire window 3 mm thick for containment of the 40 Torr gas and finally into the proportional counter itself. Single photoelectrons were predominantly generated on the copper cathode of the proportional counter chamber 1.27 cm away from the anode wire. As seen in Fig. 2 the signal from the photodiode was sent to an amplifier (Ortec-855) with a shaping time constant of 3 μs, delayed through a Quad/Generator between the gate, whose width was ~300 μs (to coincide roughly with the SE ionization signals) and then sent to a coincidence circuit (CC). The SE ionization signals were sent to the charge sensitive preamplifier, then to the amplifier (Ortec-855) with a gain of 140. The SE signals were observed on an oscilloscope and stored in the DAQ computer. A typical photodiode signal recorded in coincidence with a SE ionization signal is shown in Fig.3.

The generation of SE events in this experiment was governed, by design, by Poisson statistics [12-15], where the probability of observing $n$ events is,

$$P(n,\bar{n}) = \frac{\bar{n}^n}{n!} e^{-\bar{n}} \quad (3)$$

and where $\bar{n}$ is the average of observed events. The average number of SE and double electron (DE) events that are generated is,

$$\bar{N}_1 = P(1,\bar{n})N \quad (4)$$

and

$$\bar{N}_2 = P(2,\bar{n})N \quad (5)$$

where $N$ is the number of trials (number of the flashtube flashes). Using equations (4) and (5) the following equation is obtained,

132 $$\frac{\overline{N}_1}{\overline{N}_2} = \frac{P(1,\bar{n})}{P(2,\bar{n})} = \frac{2}{\bar{n}} \qquad (6)$$

133 In order to ensure that the average number of DE events is small compared with
134 average number of SE events it is important to measure $\bar{n}$.

135 $$\frac{\overline{N}_0}{N} = P(0,\bar{n}) = e^{-\bar{n}} \sim 1 - \bar{n} \qquad (7)$$

136 for small $\bar{n}$

137 $$\bar{n} = 1 - \frac{\overline{N}_0}{N} = 1 - \frac{R_0}{R} \qquad (8)$$

138 where $R_0$ and $R$ are the rates of zero photoelectron events and of xenon flash tube
139 flashes. As it was pointed out earlier, the flashtube was operated at a rate of $R_{flashtube}=100$
140 Hz. The rate at which zero photoelectrons were being generated is estimated with

141 $$R_0 = R_{flashtube} - (R_{obs} - R_{backgr}) \qquad (9)$$

142 where $R_{obs}$ is the observed rate of triggers and $R_{backgr}$ is the background trigger rate.
143 Therefore, from (8) and (9),

144 $$\bar{n} = 1 - \frac{R_{flashtube} - (R_{obs} - R_{backgr})}{R_{flashtube}} = \frac{R_{obs} - R_{backgr}}{R_{flashtube}} \qquad (10)$$

145 The difference in rate between SE events and background events was measured with a
146 counter (Ortec-871). The worst case in the measurements of SE events was one DE event
147 per 120 SE events during one run measurement indicating that DE events contaminated
148 the sample at < 1% for all measurements.

149

150 *E*. Gas system

151  Gas mixtures were prepared by introducing gases one by one into the evacuated
152  stainless steel vacuum vessel with the proportional counter. The initial purity of $CS_2$
153  liquid was ≥99.9% and ≥99.999% for all gas admixtures ($CF_4$, Ar, Ne and He). $CS_2$ vapor
154  was evolved from a liquid source. All experiments were done at a presurre of 40 Torr.
155  The chamber pressure was measured with a capacitance manometer (MKS, Baratron)
156  providing an accurate (0.1 Torr) determination of pressure. Measurements taken before
157  and during the experiment imply that the impurity of the gas was ≤$10^{-1}$ Torr out of 40
158  Torr in all cases.

159

160  *F*. Analysis, results and discussion

161  In analysis, several simple cuts were applied to the SE data to ensure that the
162  properties of the events could be properly measured. Specifically events lasting longer
163  than 500 μS and events clipping the digitizers were cut as most likely being due to sparks
164  or alphas. Two other cuts were placed on the Root Mean Square in Time (RMST) of the
165  SE events to ensure that they were of a width consistent with the shaping time of the
166  amplifiers and on the time difference ($\Delta T$) between the photodiode pulse and the SE
167  event to select events with consistent drift times. These latter cuts are illustrated in
168  Fig.4a. All background events are illustrated in Fig.4b, which in turn were subtracted
169  from the SE data events. To account for threshold effects, all SE spectra were fit to a
170  Polya distribution function,

171 $$P(X) = aX^b e^{-cX} \qquad (11)$$

172  where $X$-represents the channel number of a measured pulse-height distribution. The
173  mean pulse height $X_{SE}$ can be determined from the parameters $P(X)$ according to

174 $$X_{SE} = \frac{1+b}{c} \qquad (12)$$

175 where $b$ and $c$ are variables obtained from the fit. A typical SE spectrum measured in

176 pure $CS_2$ at a pressure of 40 Torr is shown in Fig. 5 with the Polya distribution fit

177 between the vertical lines.

178     Similar simple analysis cuts were applied to the $^{55}$Fe data. The measurement of an

179 average pulse-height is not meaningful for $^{55}$Fe events, however, because negative ion

180 drift spreads the ionization in time much more than the shaping time ~3 µs of the linear

181 amplifier (Ortec-855). In other words, the pulse height for $^{55}$Fe events has little to do

182 with the charge falling on the wire. The voltage integral, or area, of an event, $\Sigma$, however,

183 is proportional to the charge [7]. The mean area of an ionization signal from the $^{55}$Fe

184 source was determined with a Gaussian fit. A typical ionization signal peak from the $^{55}$Fe

185 source measured in pure $CS_2$ at 40 Torr of pressure is shown in Fig.6a. The K-escape

186 peak is also partially observed in Fig.6a. The background events were taken after the

187 measurements of ionization signals from the $^{55}$Fe source. A typical background

188 measurement taken between the measurements of the $^{55}$Fe events is shown in Fig.6b. The

189 mean areas ($\Sigma$) for $^{55}$Fe events were converted to mean pulse heights, $X_{55Fe}$, using the

190 linear relationship between the pulse height (PH) of a SE pulse and its area as shown in

191 Fig.7. Measurements of ionization yield from the $^{55}$Fe source in pure $CS_2$ and in $CS_2$ gas

192 mixtures were taken before and after SE electron measurements to ensure gain stability

193 during the experiment. A typical illustration of the gain stability in pure $CS_2$ gas during

194 the measurements is shown in Fig.8.

195     Having obtained the mean pulse height $X_{SE}$ from (12) and $X_{55Fe}$ from above, W-

196 values were calculated for pure $CS_2$ and for $CS_2$ gas mixtures using equations (1) and (2).

All results on measurements of ionization yield and W-value for the gases obtained in this work are presented in Table 1.

There are a few reported measurements of W-value in pure vapor $CS_2$ gas. The results by T. Nakayama et al. [16,17], who used Bragg-Gray principle on measuring ionization curves in pure $CS_2$ vapor at a pressure range of ~(25-150) Torr, revealed that W-value for electrons using a $^{60}Co$ source and for alpha particles of the mean energy of 2.82 MeV are 24.7±0.7 eV and 26.0±0.5 eV, respectively. The W-value for $CS_2$ gas obtained in this work agrees to within statistical errors with W-value measured for electrons in [16].

The W-values in $CS_2$ gas mixtures ($CF_4$, Ar, Ne, He), measured for the first time here to the best of our knowledge, were found to be lower than in pure $CS_2$ and are presented in Table 1. It should be pointed out that $CF_4$, Ar, Ne and He are prominent scintillating gases and the additional formation of electrons and, as a consequence, decrease of W-values is not associated with interactions of photons with the proportional counter made out of copper, which has a work function of $\varphi\sim 4.7$ eV, due to high photoabsorption and photoionization cross-sections $\sigma\sim 10^{-16}$-$10^{-17}$cm$^2$ of the photons in the wavelength range of ~100-265 nm with $CS_2$ molecules [18-20]. A detail analysis of W-values in pure $CS_2$ and in $CS_2$ with gas mixtures with further explanation will be published elsewhere [21].

## 3. Measurements of mobility

Mobility measurements were made in pure gaseous $CS_2$ and in $CS_2$ with $CF_4$, Ar, Ne and He gas mixtures at a pressure of 40 Torr. The mobility is defined with the following equation,

219 $$v = \frac{\mu E}{p} \quad (13)$$

220 where $v$ is the drift velocity, $\mu$ the mobility, $E$ is the electric field and $p$ is the gas

221 pressure [22].

222 The drift time, $\Delta t$, of SE negative ions drifting from the cathode, at radius $a$, in the

223 proportional counter to the anode wire, of radius $b$, is given by the following equation,

224 $$\Delta t = \frac{p \ln(\frac{b}{a})}{2\mu \Delta V}(b^2 - a^2) \quad (14)$$

225 where $\Delta V$ is the potential difference from cathode to anode. Measurement of the time

226 delay between the flashlamp pulse, controlled with the photodiode, with near

227 simultaneous generation of photoelectrons on the copper cathode, and the appearance of

228 an avalanche on the anode (see Fig.3.) allows one to calculate mobility.

229 $$\mu = \frac{p \ln(\frac{b}{a})}{2\Delta t \Delta V}(b^2 - a^2) \quad (15)$$

230 The data on mobility of negative ions in pure $CS_2$ and $CS_2$ gas mixtures are shown in

231 Table 2. Measurements of mobility of negative ions in pure gaseous $CS_2$ and in $CS_2$ with

232 Ar mixture in work [4] have shown that the mobility is $(5.22\pm0.90)\times10^{-5}$ atm×m$^2$/Vs,

233 which is consistent to within statistical errors with the mobility for pure $CS_2$ obtained in

234 the present work $(5.42\pm0.10)\times10^{-5}$ atm×m$^2$/Vs. Mobility in $CS_2$-Ar (90%-10%) gaseous

235 mixture at a pressure of 40 Torr measured in [4] was $(5.70\pm0.40)\times10^{-5}$ atm×m$^2$/Vs, which

236 agrees with the mobility $\sim(5.90\pm0.10)\times10^{-5})$ in $CS_2$-Ar (35-5) Torr gas mixture obtained

237 in this work. Mobility of negative ions is slightly higher in the mixtures of $CS_2$ with other gases due to repulsion between $CS_2^-$ negative ions and foreign atomic or molecular admixtures. The exchange interaction between the negative ions and the foreign admixtures leads to reduction in the effective cross-section for scattering of the ions and consequently, to an increase of the mobility of negative ions [23].

## 4. Gain measurements

Gas gain measurements from the $^{55}$Fe source were made in pure $CS_2$ gas and in $CS_2$ gas mixtures as a function of the applied voltage and are presented in Table 3. Figure 9 shows gas gain as a function of applied voltage in pure $CS_2$ and in $CS_2$ gas mixtures. Gas gain for pure $CS_2$ was calculated using the average PH of the single electron falling on the wire determined from the fit in Fig.5 and equation (12), using the gain sensitivity of the charge preamplifier (Amptek-250) ~0.16 uV/electron and taking into account the gain (140) of the amplifier (Ortec-855) for SE events. It is seen from the plot that the ionization yield increases on average by a factor of 2.3 in $CS_2$ gas mixtures at 1550 V relatively to the ionization yield in pure $CS_2$ and tends to be stable over the whole range of voltages at 1500 V and 1450 V.

## 5. Conclusion

W-value, mobility and gas gain measurements were made in $CS_2$ and $CS_2$-$CF_4$, $CS_2$-Ar, $CS_2$-Ne and $CS_2$-He gas mixtures making use of a proportional counter at 1600 V and 1550 V. The measurements show that the W-value in pure $CS_2$ obtained in this work agrees well to within statistical errors with the W-value in pure $CS_2$ for electrons in work [16]. However, the W-values in $CS_2$ with gas mixtures were found to be substantially lower than in pure $CS_2$. The mobility of negative ions in pure $CS_2$ was found to be consistent with the value obtained earlier in previous work [4]. The gas gain increases in $CS_2$ gas mixtures on the average by factor of 2.3 at 1550 V relatively to pure $CS_2$ and tends to be stable over the whole range of voltages at 1500 V and 1450 V.

## Acknowledgements

We acknowledge the support of the U.S. National Science Foundation (NSF) under grant number 0600840. Any opinions, findings, and conclusions or recommendations expressed in this material are those of the authors and do not necessarily reflect the views of the NSF.

**Figures**

Fig.1. Schematic view of the experimental set-up and electronics for measurements of 5.9 keV X-rays from the $^{55}$Fe source.

Fig.2. Schematic view of the experimental set-up and electronics for measurements of SE events using a xenon flashtube.

Fig.3. A typical SE signal recorded in coincidence with a PD signal.

Fig.4a. A plot of the Root Mean Square Time (RMST) of SE ionization signals vs the time difference ($\Delta T$) between the photodiode pulse and the SE event for a SE run.

Fig.4b. A plot of the Root Mean Square Time (RMST) of SE ionization signals vs the time difference ($\Delta T$) between the photodiode pulse and the SE event for a background run with the same live time as for the SE run

Fig.5. A SE spectrum measured in pure $CS_2$ at a pressure of 40 Torr with the Polya distribution fit between the vertical lines.

Fig.6a. A Gaussian distribution for the ionization signal measured in pure $CS_2$ at a pressure of 40 Torr with live time of ~ 48 s.

Fig.6b. Measurements of background events measured between the measurements of ionization signals from the $^{55}$Fe source in pure $CS_2$ at a pressure of 40 Torr with live time of ~51 s.

Fig.7. Pulse height vs area for SE events measured in pure $CS_2$ at a pressure of 40 Torr.

Fig.8. Gain stability vs number of runs measured in pure $CS_2$ at 40 Torr of pressure.

Fig.9. Gas gain measurements in pure $CS_2$ and in $CS_2$ gas mixtures as a function of high voltage.

Figure 1

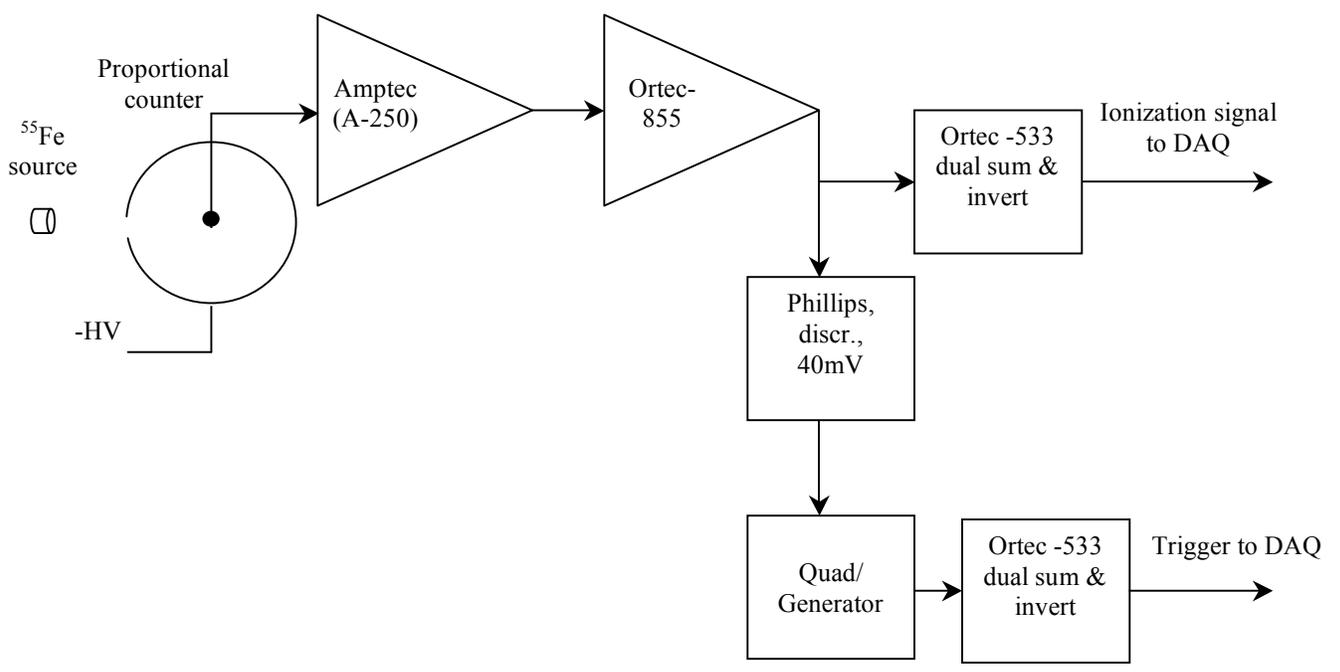

Schematic view of the experimental set-up and electronics for measurements of 5.9 keV X-rays from the $^{55}$Fe source.

Figure 2

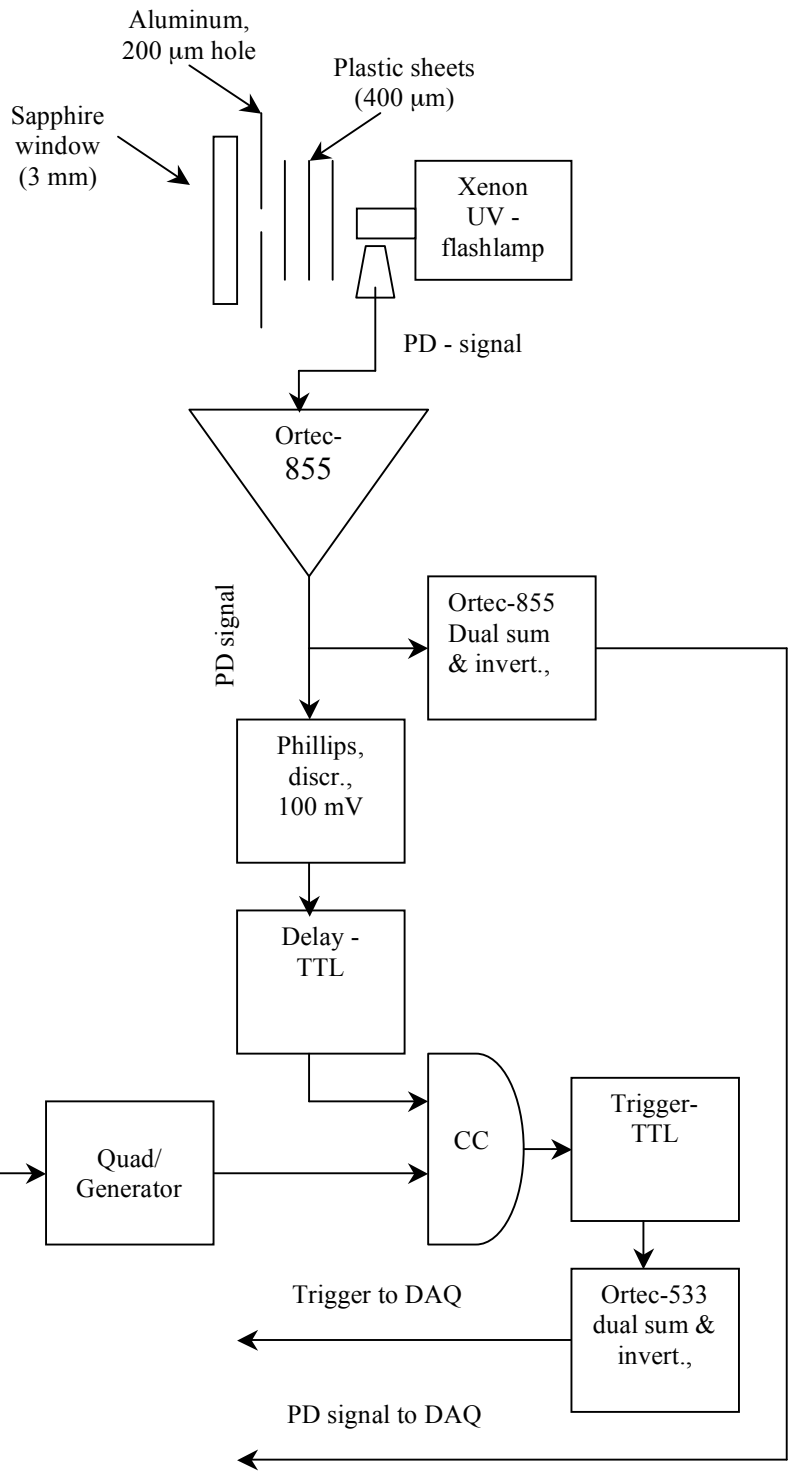

Schematic view of the experimental set-up and electronics for measurements of SE events using a xenon flashtube

Figure 3

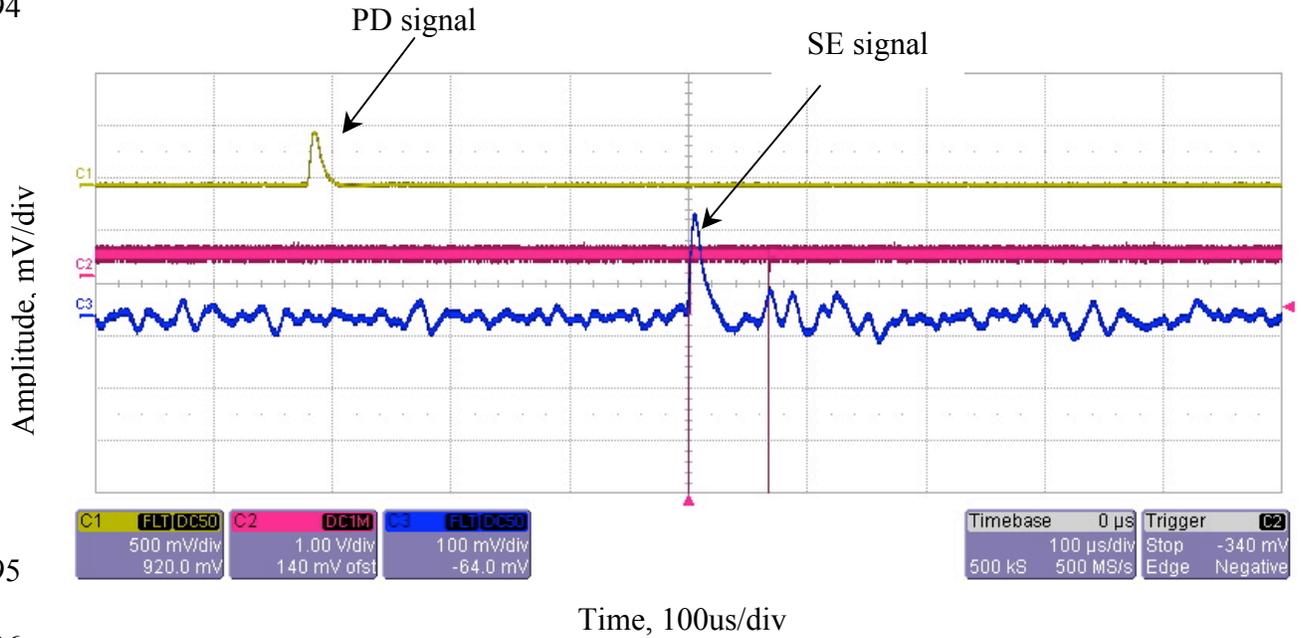

A typical SE signal recorded in coincidence with a PD signal.

406     Figure 4 a

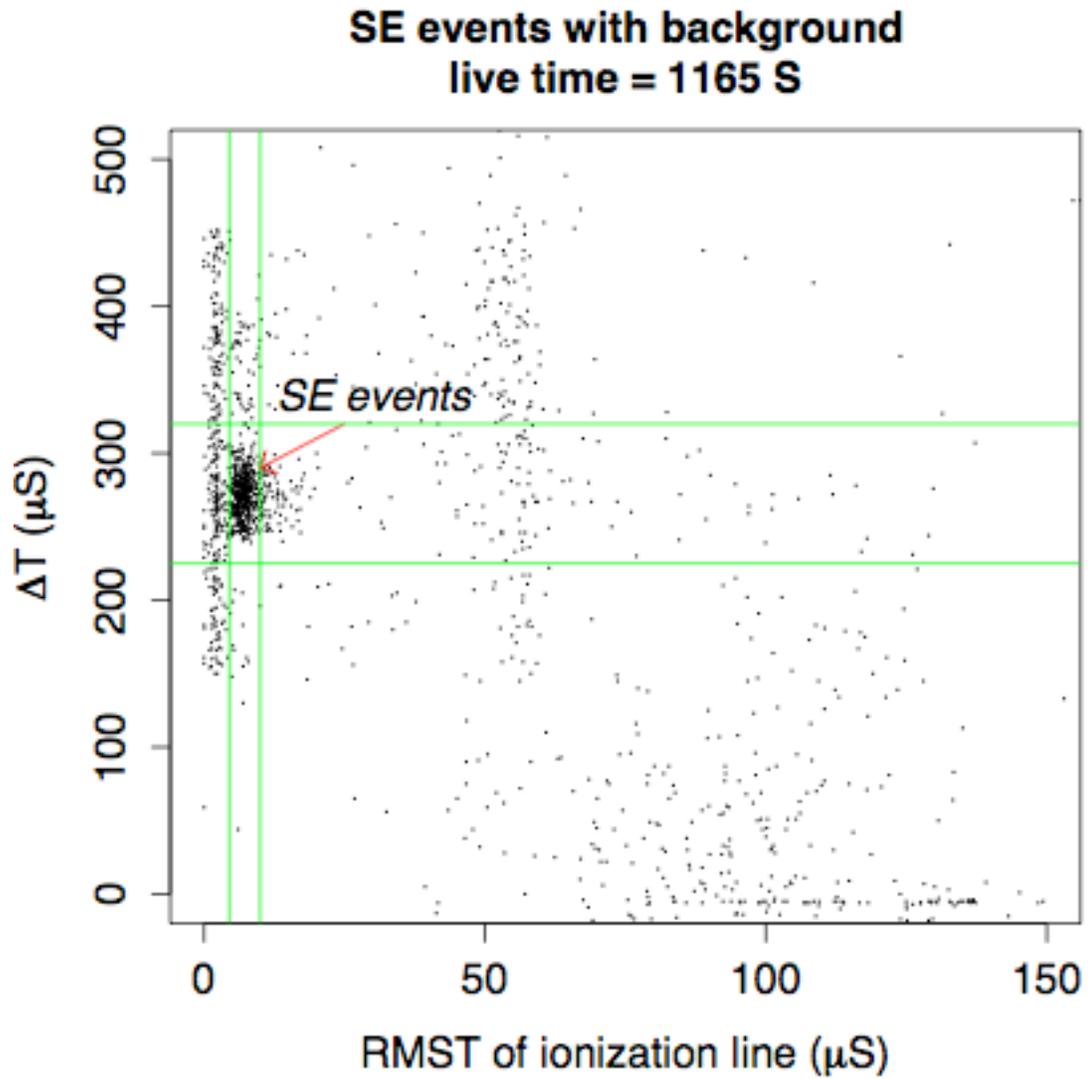

407

408

409     A plot of the Root Mean Square Time (RMST) of SE ionization signals vs the time

410     difference ($\Delta T$) between the photodiode pulse and the SE event for a SE run.

411

412

413

414

Figure 4 b

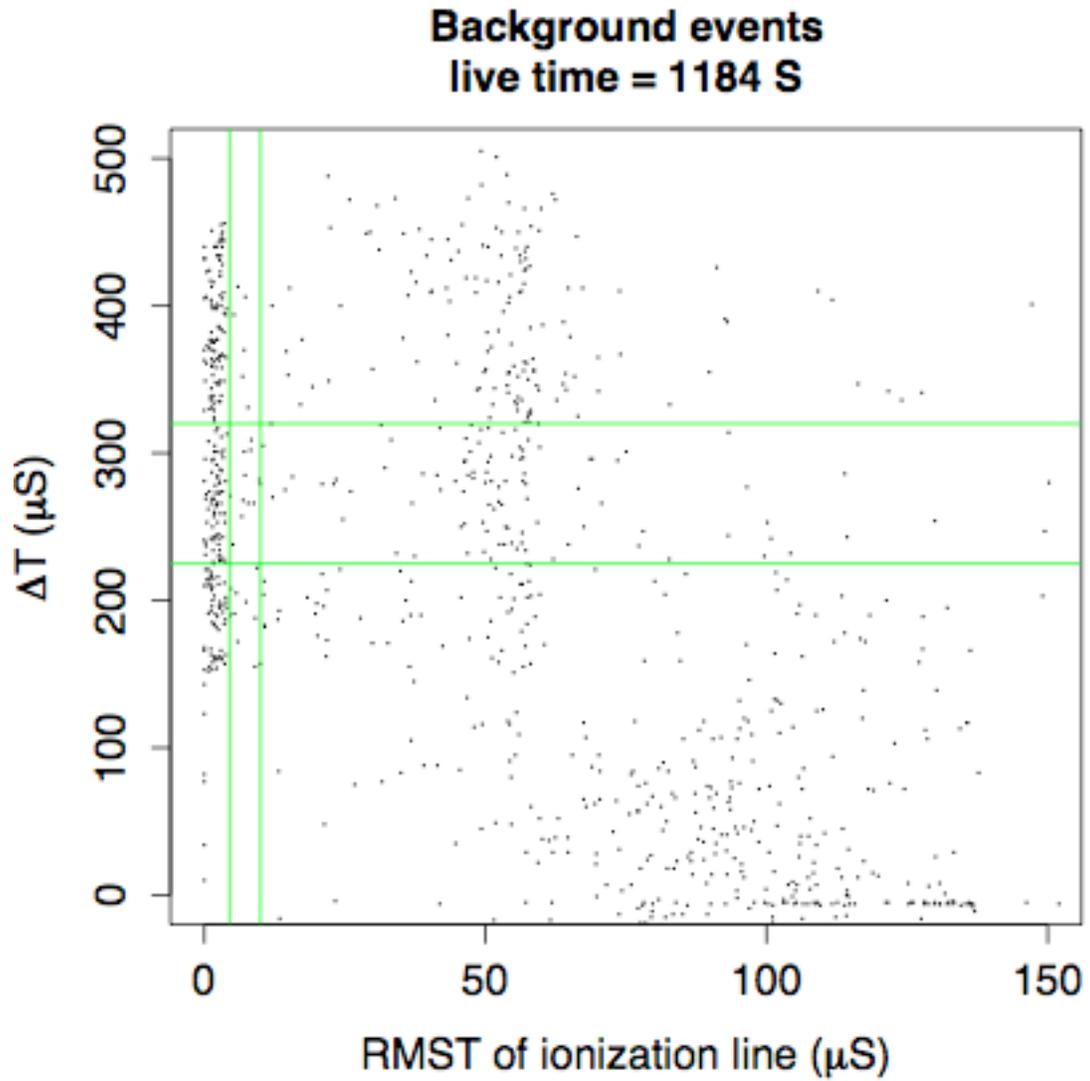

A plot of the Root Mean Square Time (RMST) of SE ionization signals vs the time difference (*ΔT*) between the photodiode pulse and the SE event for a background run with the same live time as for the SE run

424    Figure 5

425

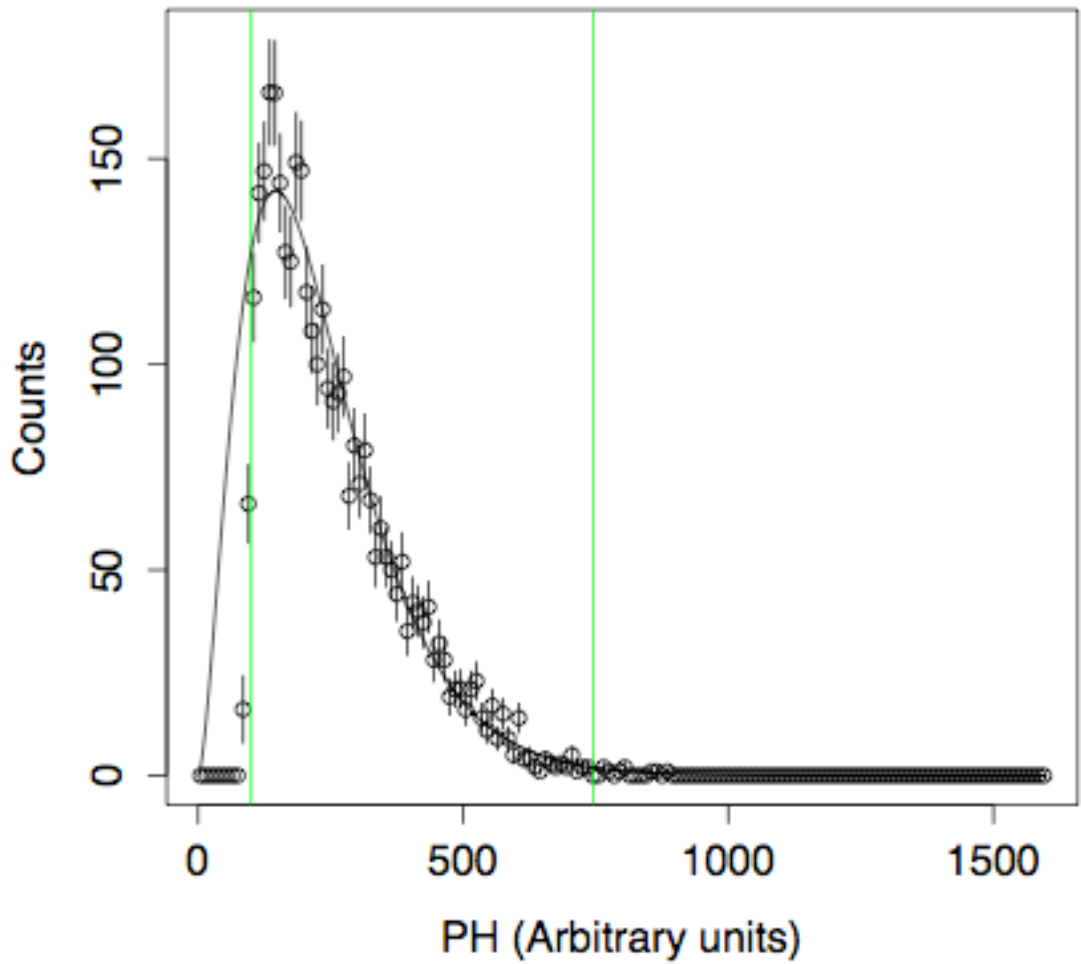

426

427

428    A SE spectrum measured in pure $CS_2$ at a pressure of 40 Torr with the Polya

429    distribution fit between the vertical lines.

430

431

432

433

434    Figure 6a

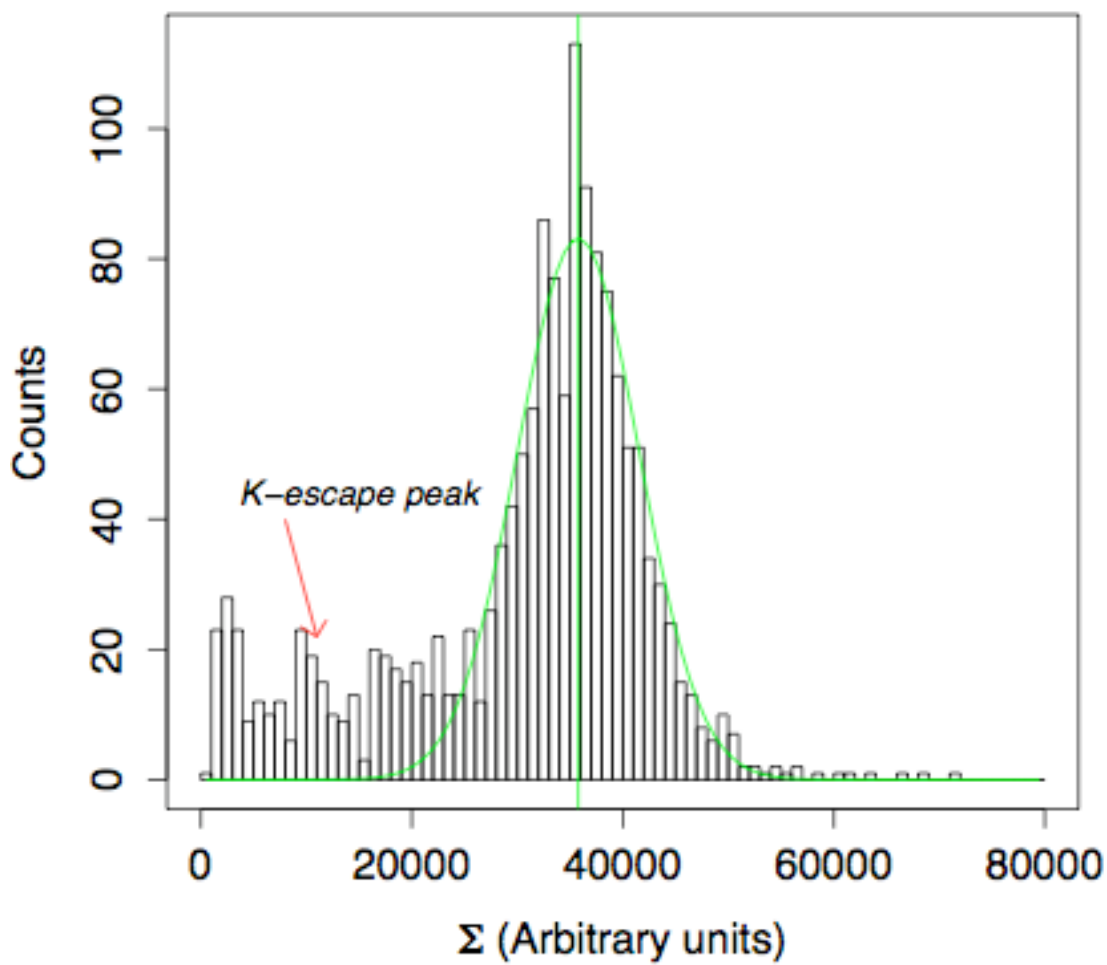

A Gaussian distribution for the ionization signal measured in pure $CS_2$ at a pressure of 40 Torr with live time of ~ 48 s.

443  Figure 6b

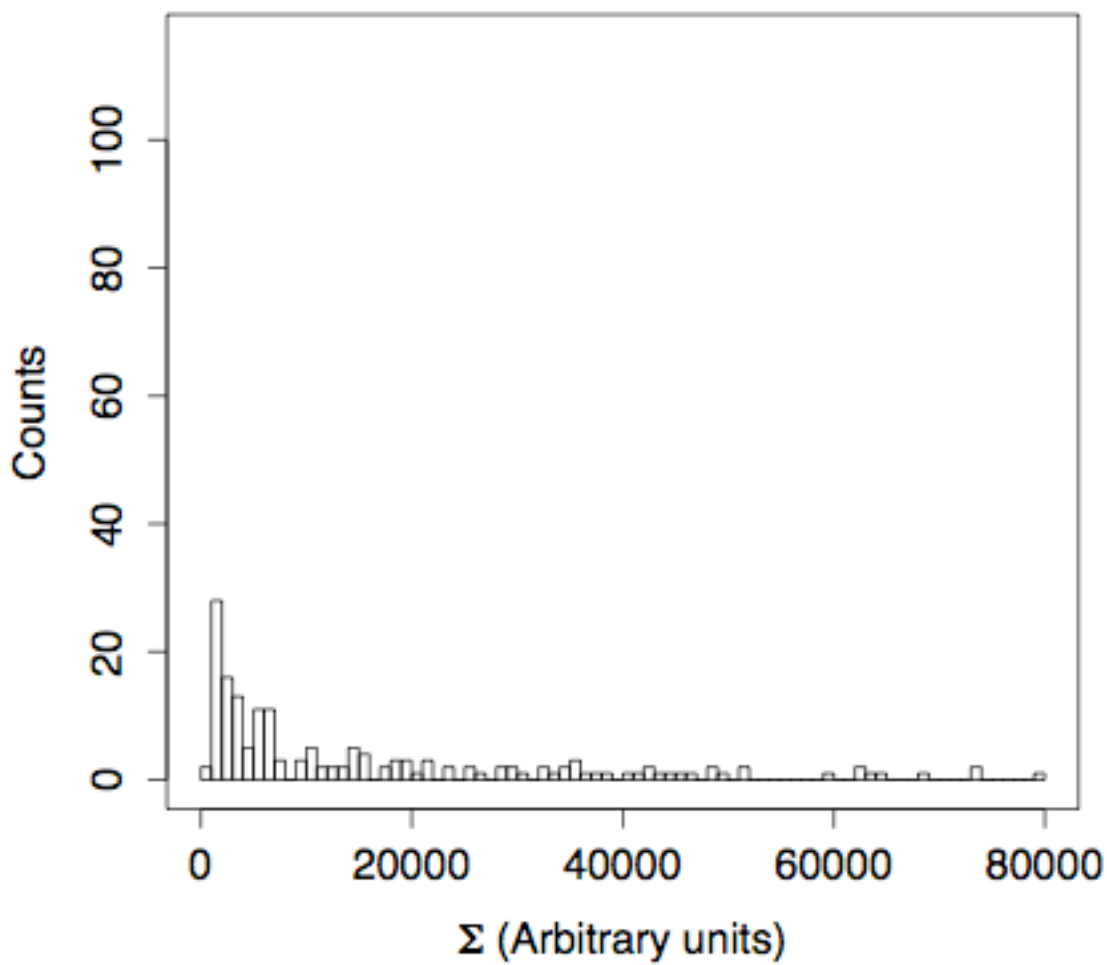

Measurements of background events measured between the measurements of ionization signals from the $^{55}$Fe source in pure $CS_2$ at a pressure of 40 Torr with live time of ~51 s.

452  Figure 7

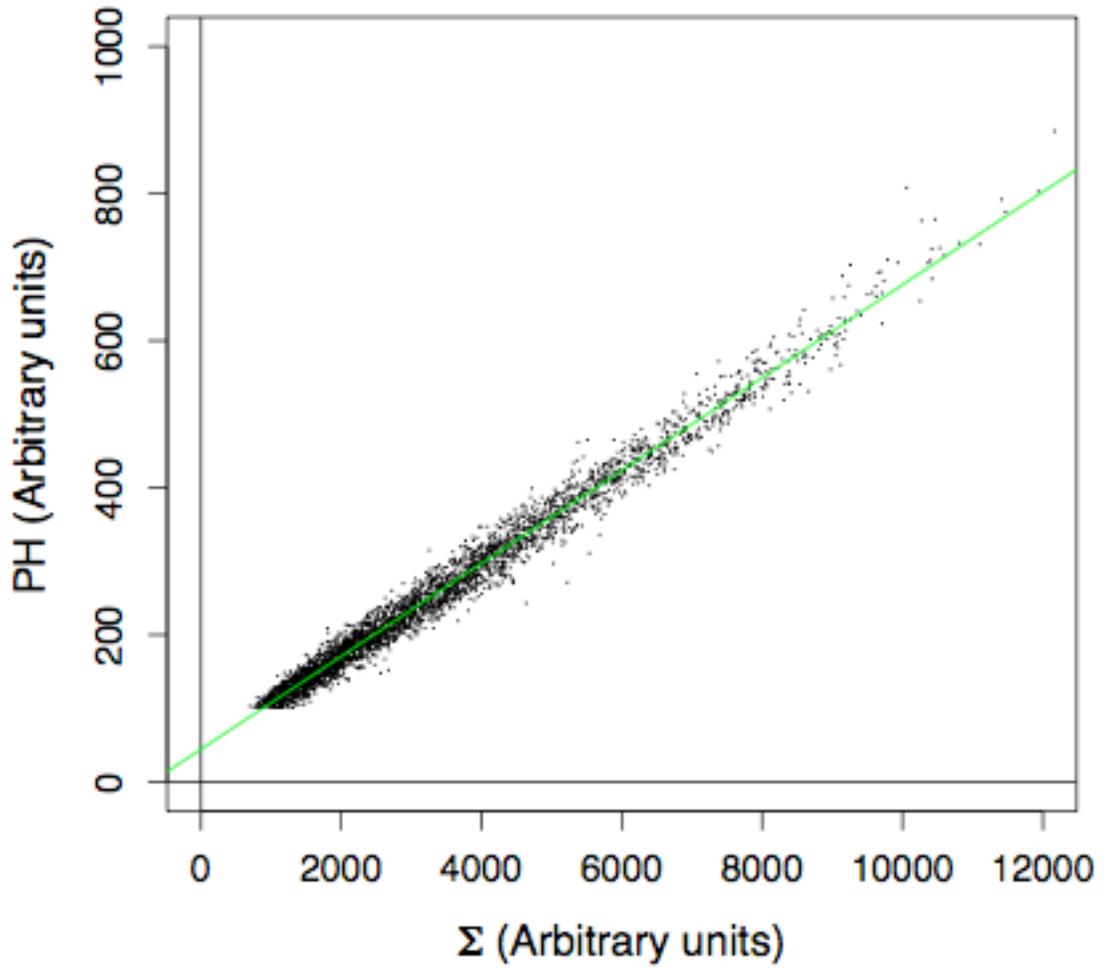

455  Pulse height vs area for SE events measured in pure $CS_2$ at a pressure of 40 Torr.

461     Figure 8

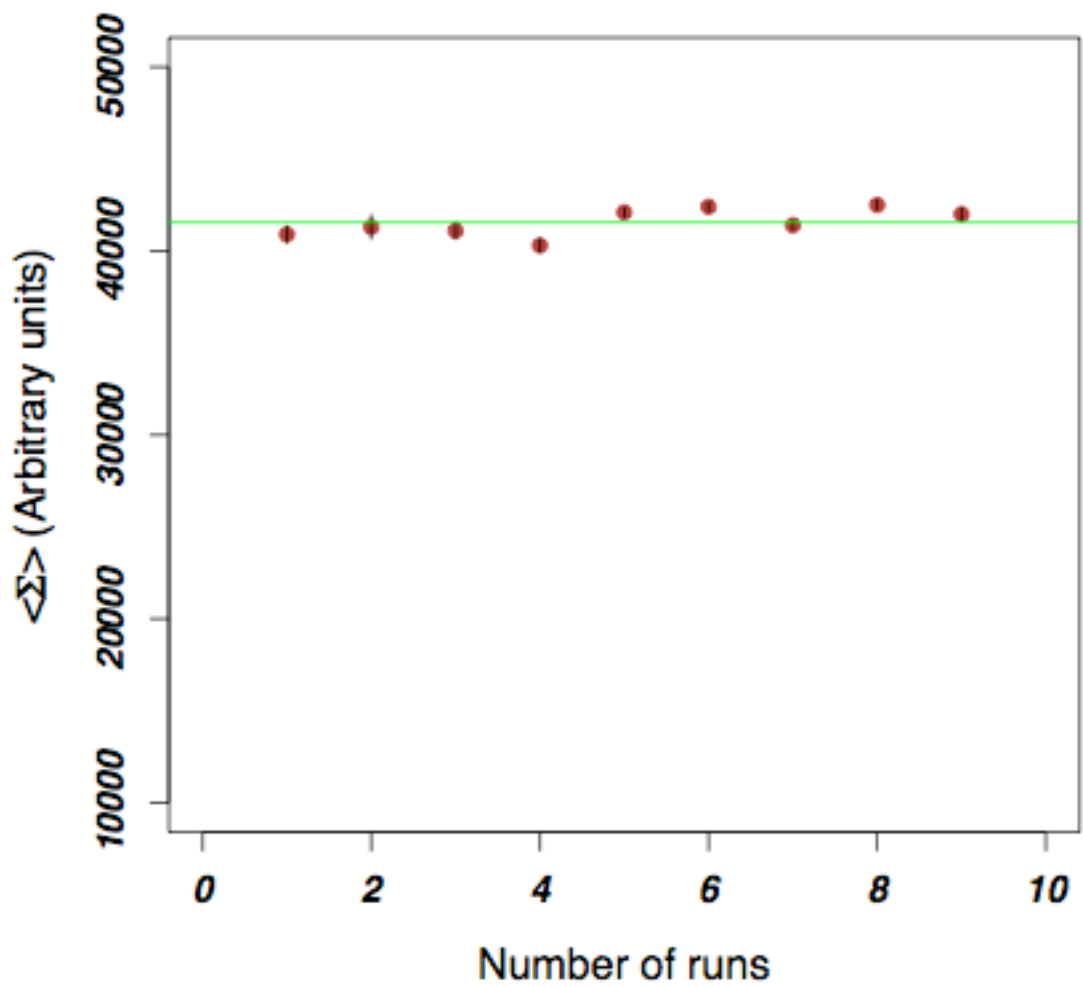

462

463     Gain stability vs number of runs measured in pure CS$_2$ at 40 Torr of pressure.

464

465

466

467

468

469

470                                                                                    Table 1

| Gases, (Torr) | Voltage, V | Ionization yield | W-value, eV | W-value, eV Pure $CS_2$ (other works) |
|---|---|---|---|---|
| Pure $CS_2$, (40) | 1600 | 280±36 | 21.1±2.7(stat)±3(syst) | 24.7±0.7[16] 26.0±0.5 [17] |
| $CS_2$-$CF_4$, (30-10) | 1550 | 360±40 | 16.4±1.8(stat)±2(syst) | |
| $CS_2$-Ar, (35-5) | 1550 | 450±50 | 13.1±1.5(stat)±2(syst) | |
| $CS_2$-He, (35-5) | 1550 | 340±60 | 17.3±3.0(stat)±3(syst) | |
| $CS_2$-Ne, (35-5) | 1550 | 360±70 | 16.3±3.0(stat)±3(syst) | |

471

472

473

474

475

476

477

478

479

480

481

482

483

484 Table 2

| Gases, Torr | Voltage, V | Drift time, us | Mobility factor, $\mu$, $\frac{atm \times m^2}{Vs}$ | Mobility factor, $\mu$, $\frac{atm \times m^2}{Vs}$ (other works) |
|---|---|---|---|---|
| Pure CS$_2$, (40) | 1600 | 270.8±0.2 | (5.42±0.10)×10$^{-5}$ | (5.22±0.90)×10$^{-5}$ [4] |
| CS$_2$-CF$_4$, (30-10) | 1550 | 250.1±0.2 | (6.10±0.10)×10$^{-5}$ | |
| CS$_2$-Ar, (35-5) | 1550 | 257.4±0.2 | (5.90±0.10)×10$^{-5}$ | (5.70±0.40)×10$^{-5}$ [4] |
| CS$_2$-He, (35-5) | 1550 | 252.0±0.3 | (6.02±0.10)×10$^{-5}$ | |
| CS$_2$-Ne, (35-5) | 1550 | 248.2±0.3 | (6.12±0.10)×10$^{-5}$ | |

485

486

487

488

489

490

491

492

493

494

495

496

| Gases, Torr | Voltage, V | Gas Gain |
|---|---|---|
| Pure $CS_2$, (40) | 1600 | 4910±640 |
|  | 1550 | 2460±340 |
|  | 1500 | 1230±200 |
|  | 1450 | 610±110 |
|  | 1400 | 380±80 |
| $CS_2$-$CF_4$, (30-10) | 1550 | 5800±700 |
|  | 1500 | 2760±360 |
|  | 1450 | 1320±170 |
|  | 1400 | 670±100 |
|  | 1350 | 320±80 |
| $CS_2$-He, (35-5) | 1550 | 4910±900 |
|  | 1500 | 2460±440 |
|  | 1450 | 1230±230 |
|  | 1400 | 590±110 |
|  | 1350 | 270±90 |
| $CS_2$-Ne, (35-5) | 1550 | 5800±750 |
|  | 1500 | 2760±360 |
|  | 1450 | 1260±180 |
|  | 1400 | 640±110 |
|  | 1350 | 260±110 |
| $CS_2$-Ar, (35-5) | 1550 | 5360±910 |
|  | 1500 | 2230±380 |
|  | 1450 | 450±90 |

498

499

500

501

502

503

504

505

506  Figure 9

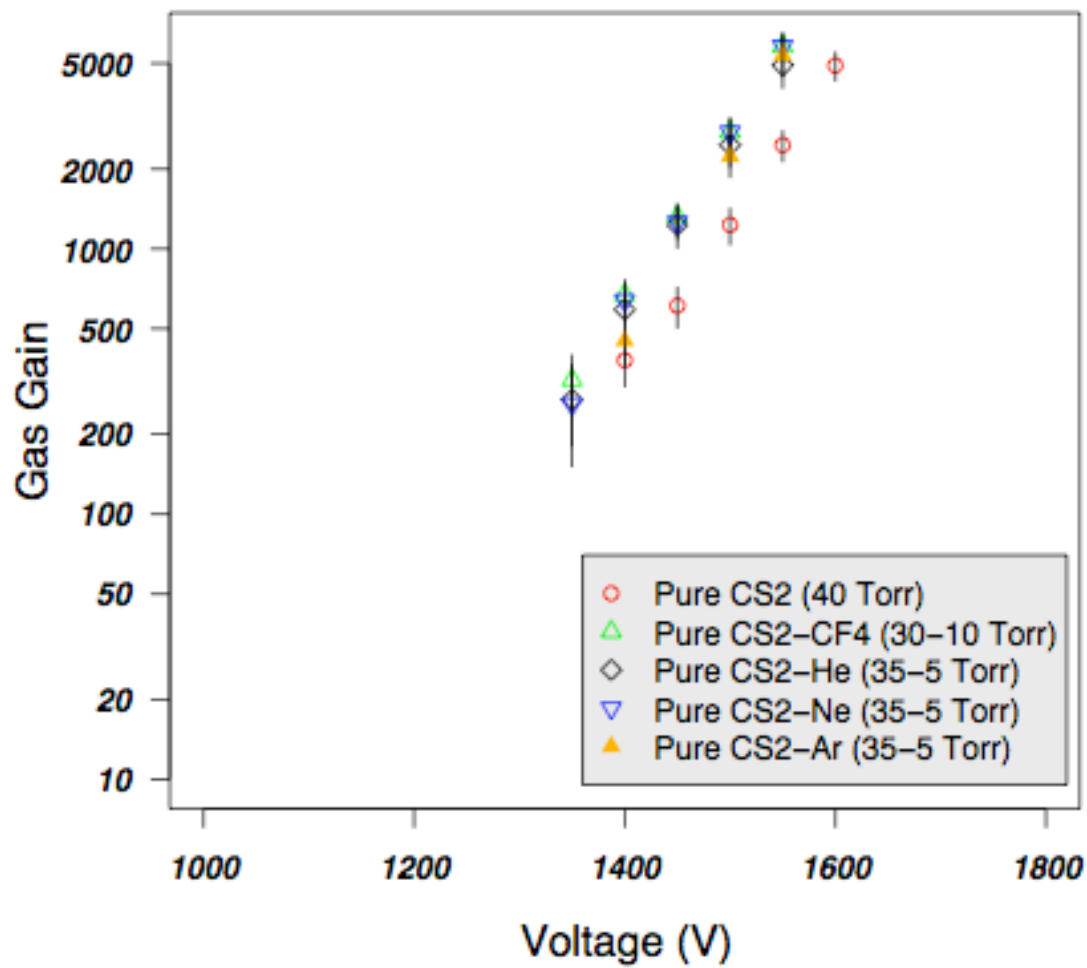

508  Gas gain measurements in pure $CS_2$ and in $CS_2$ gas mixtures as a function of high
509  voltage.